\documentstyle[12pt,epsfig,axodraw]{article}
\newcommand{\e}{{\rm  e}}

\newcommand{\beq}{ \begin{eqnarray} }
\newcommand{\eeq}{ \end{eqnarray} }
\newcommand{\beqstar}{ \begin{eqnarray*} }
\newcommand{\eeqstar}{ \end{eqnarray*} }

\begin{document}

\begin{titlepage}

\begin{center}

\hfill KEK-TH-577\\
\hfill \today

  {\large  
	Lepton-Flavor Violation in Supersymmetric Models\footnote{
Talk presented at International Workshop on JHF Science (JHF98), March
4-7, 1998, KEK, Tsukuba, Japan.
}

}
  \vskip 0.5in {\large
    J.~Hisano
	}
\vskip 0.4cm 
{\it 
Theory Group, KEK, Oho 1-1, Tsukuba, Ibaraki 305-0801, Japan
}
\vskip 0.5in

\abstract {
Theoretical aspects of lepton-flavor violating (LFV) processes on the 
supersymmetric models are reviewed.  In particular, we show that, 
assuming the minimal supergravity scenario, the LFV interaction 
at the higher energy scale leads to the the LFV processes, which 
may be accessible in near future experiments.
}
\end{center}
\end{titlepage}

\setcounter{footnote}{0}

\section{Introduction}

It is well-known that the standard model (SM) has three kinds of conserved 
quantities, the baryon and the lepton numbers, and the lepton-flavor numbers. 
While search for violation of the conservation lows is still a powerful 
way to prove physics beyond the standard model, the processes 
are rare  since they are induced by the nonrenormalizable operators 
suppressed by powers of the energy scale beyond the standard model. 

The minimal supersymmetric (SUSY) extension of the standard model (MSSM),
that is motivated as a solution of the gauge hierarchy problem,
is one of the most promising model beyond the standard model.
In this model the lepton flavor violation (LFV) is considered as 
one of the most important prediction \cite{EN}. Supersymmetry is a symmetry 
between bosons and 
fermions. If this symmetry was exact, leptons and the superpartners,
called as slepton, had a common mass terms, and then the lepton flavor was 
conserved, identical to the standard model. However, supersymmetry has to be 
violating at low energy, since superpartners for the SM particles are not still
discovered at all. Then, associated with the breaking of supersymmetry, the 
lepton flavor may be violating through the SUSY breaking mass terms for sleptons.
Since the SUSY breaking scale should be below $O(1)$TeV from a point of 
naturalness, the LFV processes, such as $\mu\rightarrow \e \gamma$, are 
expected to be observed in near future experiments as a signature of 
supersymmetry. In fact, the present experimental upper bounds on the event 
rates have already given a constraint on this model.

The predicted event rates of the LFV processes in this model depend on the 
detail of the slepton mass matrixes. They are determined by both the mechanism
of generation of SUSY breaking terms in the MSSM and the physics 
beyond the MSSM.

Nowadays, two kinds of the models to generate the SUSY breaking terms in 
the MSSM are considered. One is the minimal supergravity scenario \cite{nilles}
and other is the gauge mediated SUSY breaking scenario \cite{RG}. These are 
proposed in order to suppress the FCNC processes. As we mentioned above, 
the experiments to search for the LFV processes have already given 
constraints on the slepton mass matrixes. Also, arbitrary squark mass 
matrixes lead to excesses of $K^0-\overline{K}^0$ mixing, 
$b\rightarrow s \gamma$ event rate, and so on.
In order to suppress these processes, the SUSY breaking mass terms for squarks 
and sleptons have to be almost flavor-independent. This can be derived if 
the SUSY breaking terms in the MSSM are generated  by mediation of a
flavor-independent interaction from a sector where supersymmetry is 
spontaneous broken. The candidates of the mediator are gravity in the 
minimal supergravity scenario, and gauge interactions
in the gauge mediated SUSY breaking scenario.

In these scenarios, the LFV processes can be suppressed below the present 
experimental bounds. However, we still have an interesting possibility 
to observe the LFV processes in near future experiments. Besides the MSSM,
several models with the LFV interaction are proposed. For example, the grand 
unified theories (GUT's) \cite{PL}, the seesaw mechanism with the right-handed 
neutrinos \cite{seesaw}, and so on. In the 
non-supersymmetric model, the event rates of the LFV processes are suppressed 
by powers of the the energy scale ($M_{\rm LFV}$), and they are too rare to 
be observed. However, if supersymmetry exists, that is not necessary valid. If 
the generic energy scale of the SUSY breaking mediators ($M_{\rm M}$) is 
larger than $M_{\rm LFV}$,  the sizable LFV SUSY breaking mass terms for
sleptons may be generated radiatively, not suppressed by powers of 
$M_{\rm LFV}$ \cite{HKR}. In the minimal supergravity scenario $M_{\rm M}$ is 
considered to be $M_{\rm G}\sim$10$^{18}$GeV. Then, if the model with LFV interaction exits 
at larger energy scale than the SUSY breaking scale, the LFV processes may be 
accessible in near future experiments. In fact, some models predict the 
branching ratio of $\mu\rightarrow \e\gamma$ at most one or two 
orders of magnitude below the experimental upper bound. 

In a representative model of the gauge mediated SUSY breaking scenario \cite{DNNS}
$M_{\rm M}$ is 10$^{(4-5)}$GeV, while the other models exist. 
In this case, the LFV mass terms for sleptons, generated by the LFV interaction 
at larger energy scale than $M_{\rm M}$, are suppressed by powers of  
$M_{\rm M}/M_{\rm LFV}$. Then, the LFV processes are too rare, 
similar to the non-supersymmetric case \cite{DS}.

In this article, we review the LFV processes in the supersymmetric models,
assuming the minimal supergravity scenario, and show that near future 
experiments may observe them. In next section, 
we will explain how the LFV slepton masses are generated by the radiative 
correction. In section 3 we will show the branching ratio of $\mu\rightarrow 
\e \gamma$ in the typical models. Section 4 is conclusion and discussion.
Here, the other LFV processes will be discussed. 

\section{Radiative generation of LFV}

First, we introduce the MSSM, briefly. The Yukawa couplings giving masses to 
quarks and leptons in the MSSM are given by the following superpotential,
\newcommand{\mup}{m_{\tilde{u}ij}}
\newcommand{\md}{m_{\tilde{d}ij}}
\newcommand{\me}{m_{\tilde{e}ij}}
\newcommand{\mnu}{m_{\tilde{\nu}ij}}
\newcommand{\mq}{m_{\tilde{Q}ij}}
\newcommand{\ml}{m_{\tilde{L}ij}}
\newcommand{\mhd}{m_{\tilde{h1}ij}}
\newcommand{\mhu}{m_{\tilde{h2}ij}}
\newcommand{\sq}{{\tilde{q}_L}}
\newcommand{\su}{{\tilde{u}_R}}
\newcommand{\sd}{{\tilde{d}_R}}
\newcommand{\slep}{{\tilde{l}_L}}
\newcommand{\se}{{\tilde{e}_R}}
\newcommand{\snu}{{\tilde{\nu}_R}}
\begin{eqnarray}
W_{\rm MSSM}&=&
\phantom{+}f_{l_i} \overline{E}_i L_i \overline{H}_f
  + f_{d_i} Q_i \overline{D}_i \overline{H}_f
  + V_{\rm CKM}^{ji} f_{u_j} Q_i \overline{U}_{j} H_f  
\label{superpotential}
\end{eqnarray}
where $L(\equiv(N,E))$ represents a chiral multiplet of an $SU(2)_L$ doublet
lepton and  $\overline{E}$ an $SU(2)_L$ singlet charged lepton.\footnote{
In this article capital letters represent chiral superfields, and 
small letters are referred as the components.
}
Similarly, $Q(\equiv(U,D))$, $\overline{U}$ and $\overline{D}$ represent 
chiral multiplets of quarks of a $SU(2)_L$ doublet and two singlets with 
different $U(1)_Y$ charges. Three generations of leptons and quarks 
are assumed, and then the subscripts $i$ and $j$ run over 1 to 3. 
Two Higgs doublets with opposite hypercharge,  
$\overline{H}_f(\equiv(\overline{H}_f^0,\overline{H}_f^-)$)
and $H_f(\equiv(N_f^+,N_f^0)$), 
are introduced in order for the gauge anomaly by the Higgsino doublets, 
the fermionic partners of the doublet Higgs bosons, to cancel out. 
In Eq.~(\ref{superpotential}) $V_{\rm CKM}$ is the Cabibbo-Kobayashi-Maskawa (CKM)
matrix. For convenience in later discussion, the ratio of the vacuum 
expectation values of the doublet Higgs bosons is referred as 
\begin{eqnarray}
\tan\beta&\equiv&\frac{\langle h_f \rangle}{\langle \overline{h}_f \rangle}.
\end{eqnarray}

As we explained in Introduction, supersymmetry is broken in the MSSM, and 
in general the SUSY breaking terms for squarks and sleptons are given by
\begin{eqnarray}
\label{softbreaking}
-{\cal{L}}_{\rm soft}&=&(m_{\tilde q}^2)_i^j {\tilde q}^{\dagger i}
{\tilde q}_{j}
+(m_{\tilde{\bar{u}}}^2)^i_j {\tilde{\bar{u}}}_{i}^* {\tilde{\bar{u}}}^j
+(m_{\tilde{\bar{d}}}^2)^i_j {\tilde{\bar{d}}}_{i}^* {\tilde{\bar{d}}}^j
+(m_{\tilde l}^2)_i^j {\tilde l}^{\dagger i}{\tilde l}_{j}
+(m_{\tilde{\bar{e}}}^2)^i_j {\tilde{\bar{e}}}_{i}^* {\tilde{\bar{e}}}^j
\nonumber \\
& &
+(
A_d^{ij} \bar{h}_f  {\tilde{\bar{d}}}_{i}{\tilde q}_{j}
+A_u^{ij}  h_f {\tilde{\bar{u}}}_{i} {\tilde q}_{j} 
+A_l^{ij} \bar{h}_f {\tilde{\bar{e}}}_{i}{\tilde l}_{j} +h.c.).
\end{eqnarray}
Here, the terms on the first line are  soft breaking mass terms 
for  sleptons and squarks
while $A_u$, $A_d$, and  $A_l$ are the SUSY breaking parameters 
associated with the supersymmetric Yukawa couplings. The off-diagonal 
components of $(m_{\tilde{\bar{e}}}^2)$, $(m_{\tilde l}^2)$, and $A_l$ are 
lepton-flavor violating. 

In the minimal supergravity scenario, the 
SUSY breaking masses for squarks and sleptons and $A_u$, $A_d$, and $A_l$  
are given  at tree level as follows,
\begin{eqnarray}
&(m_{\tilde q}^2)_i^j=(m_{\tilde{\bar{u}}}^2)^i_j=(m_{\tilde{\bar{d}}}^2)^i_j=
(m_{\tilde l}^2)_i^j=(m_{\tilde{\bar{e}}}^2)^i_j =m_0^2 \delta^i_j,
&
\nonumber\\
&A_{u}^{ij}=a_0 m_0 f_{u}^{ij},~
A_{d}^{ij}=a_0 m_0 f_{d}^{ij},~
A_{l}^{ij}=a_0 m_0 f_{l}^{ij}.
\label{BC}
\end{eqnarray}
The universality of squark and slepton masses suppresses
the SUSY contribution to the FCNC processes. However, this 
is not stable under the quantum correction. If  fields with 
the mass larger than the SUSY breaking scale have LFV interactions, the LFV
SUSY breaking mass terms are generated radiatively. Then, assuming
the minimal supergravity scenario,  we can probe physics beyond the MSSM 
through the LFV processes. We  present it in two representative models with 
LFV interactions, the minimal SUSY SU(5) GUT and the seesaw mechanism with 
the right-handed neutrinos. 

The minimal SUSY SU(5) GUT  unifies the three gauge groups in the SM 
in order to explain the 
electric-charge quantization, and from the prediction of the gauge 
coupling unification, it has known
that the GUT scale is 10$^{16}$GeV. In this model, since quarks and leptons 
are embedded in common SU(5) multiplets, lepton flavor is violating 
associated with quarks. Especially interesting, in this model the large 
top quark Yukawa coupling enhances the LFV interaction, and 
the radiatively-induced LFV masses for sleptons are so large that 
$\mu\rightarrow \e\gamma$  may be accessible in near future experiments 
\cite{BH}.

In the minimal SUSY SU(5) GUT, both quarks and leptons are embedded in 
$\Phi({\bf 5}^*)$ and $\Psi({\bf 10})$ as follows,
\begin{eqnarray}
\Psi &=& \frac{1}{\sqrt{2}}
\left( \begin{array}{ccccc} 
  0 & \overline{U} & -\overline{U}   & U &  D \\
    &    0         &  \overline{U}   & U &  D \\
    &           &    0               & U &  D \\ 
    &           &                    &  0  &  \overline{E} \\ 
    &           &                    &     & 0  
\end{array} \right), \\
\Phi &=& \left(
\begin{array}{ccccc}
\overline{D}&\overline{D}& \overline{D}& E& -N 
\end{array}
\right),
\end{eqnarray}
where we suppress the generation indices. The Higgs doublets $H_f$ and 
$\overline{H}_f$ exit in $H({\bf 5})$ and $\overline{H}({\bf 5}^*)$ 
with Higgs triplets with SU(3)$_C$ color, $H_c$ and 
$\overline{H}_c$, as 
\begin{eqnarray}
H &=& \left(
\begin{array}{ccccc}
H_c &H_c & H_c &H^+_f & H_f^0
\end{array}
\right) , 
\nonumber\\
\overline{H}&=& \left( 
\begin{array}{ccccc}
\overline{H}_{c} &\overline{H}_{c}&\overline{H}_{c} &
\overline{H}^-_f & -\overline{H}^0_f
\end{array}
\right).
\end{eqnarray}
Since the Higgs triplets have a baryon- and lepton-number violating 
interaction leading to proton decay \cite{dimen5}, the masses should be
as large as the GUT scale at least \cite{proton}. 

In this model the superpotential 
leading to the quark and lepton masses is given as
\begin{eqnarray}
W_{\rm SU(5)} &=& \frac14 
V_{\rm CKM}^{ki} f_{u_k} \e^{i\theta_k} V_{\rm CKM}^{kj} \Psi_i \Psi_j H
     + \sqrt{2} f_{d_i} \Psi_i \Phi_{i} \overline{H} 
\label{Yukawa}
\end{eqnarray}
where $\theta_i$'s $(i=1-3)$ are additional phases which satisfy
$\theta_1+\theta_2+\theta_3 =0$. After redefinition of the SM fields
we get 
\begin{eqnarray}
W_{\rm SU(5)}&=&
\phantom{+} M_{\rm MSSM}
\nonumber \\
 & &
  + V_{\rm KM}^{ji} f_{u_j} 
    \overline{E}_i \overline{U}_j H_c 
  -  f_{d_i} Q_i L_i \overline{H}_c
\nonumber\\
 & &
  - \frac12 V_{\rm KM}^{ki} f_{u_k} \e^{i\theta_k} V^{kj}_{\rm KM} 
    Q_i Q_j H_c
  + V_{\rm KM} ^{ij\star} f_{d_j} \e^{-i\theta_i}  
    \overline{U}_i \overline{D}_j \overline{H}_c. 
\label{minimal}
\end{eqnarray}
where $f_{d_i} = f_{l_i}(i=1-3)$, which means that mass ratio of 
charged leptons and down-type quarks is predictable in this model.
In the first term on the second line the right-handed leptons have 
flavor-violating interaction with the Higgs triplet, which is 
controlled by the CKM matrix. This interaction leads to  non-negligible 
LFV masses for the right-handed sleptons and off-diagonal components
of $A_l$ through the radiative correction at one-loop level as
\begin{eqnarray}
(m_{\tilde{\bar{e}}}^2)^i_j  &=& 
-\frac{3}{8\pi^2} 
f_{u_3}^2 V_{\rm CKM}^{3i} V_{\rm CKM}^{3j\star}
(3 +a_0^2) m_0^2
\log\frac{M_{\rm G}}{M_{\rm GUT}},
\label{mer}
\\
A_l^{ij} &=& 
-\frac{9}{16\pi^2} 
\left(
f_{u_3}^2 V_{\rm CKM}^{3i} V_{\rm CKM}^{3j\star} f_l^j 
\right)a_0 m_0
\log\frac{M_{\rm G}}{M_{\rm GUT}},
\end{eqnarray}
with $i\ne j$. Here, we ignored the Yukawa coupling constants 
except for that of top quark. The prefactor 3 in Eq.~(\ref{mer}) 
comes from a color factor
in the loop diagram, and it enhances the radiative correction to the slepton 
masses with the large top quark Yukawa coupling constant.
On the other hand, the Yukawa coupling of the left-handed leptons to the 
Higgs triplet is diagonal. Though the off-diagonal components 
are induced at the higher orders, the effect on the LFV masses for the 
left-handed sleptons is negligibly small. 

It is well-known that the mass ratios of down-type quarks and charged 
leptons in the first and the second generations cannot be explained 
in the minimal SU(5) SUSY GUT, while the bottom-tau ratio is justified 
in some regions. If the extra interaction of $\Phi$ is introduced in 
order to care it, even if it is a nonrenormalizable interaction, the 
left-handed leptons may also have flavor-violating interaction. In this 
case the left-handed sleptons may get the LFV masses from the radiative 
corrections \cite{ACH}. 

Next, let us discuss a case of the seesaw mechanism with the right-handed
neutrinos. After introducing the right-handed neutrinos ($\overline{N}$) to 
the MSSM, the superpotential becomes
\begin{eqnarray}
W_{\nu_R} &=& W_{\rm MSSM} 
+ f_{\nu_i} V_{\rm LM}^{ij} \overline{N}_i L_j H_f
+ M_{\nu_R}^{ij}  \overline{N}_i\overline{N}_j.
\end{eqnarray}
The lepton has a mixing matrix $V_{\rm LM}$ even after redefinition of fields,
similar to the CKM matrix in the quark sector, and  then the lepton flavor is 
violating. By the see-saw mechanism the small neutrino masses ($m_{\nu}$) are 
induced as
\begin{eqnarray}
(m_{\nu})^{ij} &=& V_{\rm LM}^{Tik} f_{\nu_k} (M_{\nu_R}^{-1})^{kl} 
                  f_{\nu_l}  V_{\rm LM}^{lj} \langle h_f \rangle^2,
\end{eqnarray}
since $m_{\nu}$'s become zero in a limit $M_{\nu_R}\rightarrow 
\infty$. If the tau-neutrino mass ($m_{\nu_\tau}$) is  about 10eV  so that it constitutes 
the hot component of the dark matter of the Universe, 
 the right-handed neutrino mass scale is 
expected to be 10$^{(12-13)}$GeV, assuming  $f_{\nu_3}$ is as large as the 
top quark Yukawa coupling constant.

In this model the LFV mass terms for the left-handed sleptons can be induced
radiatively \cite{BM}. The LFV off-diagonal components of $(m_{\tilde l}^2)$ and $A_l$
are given at one-loop level as
\begin{eqnarray}
(m_{\tilde l}^2)^i_j  &=& 
-\frac{1}{8\pi^2} 
f_{\nu_3}^2 V_{\rm LM}^{3i} V_{\rm LM}^{3j\star}
(3 +a_0^2) m_0^2
\log\frac{M_{\rm G}}{M_{\nu_R}},
\\
A_l^{ij} &=& 
-\frac{3}{16\pi^2} 
\left(
f_l^i f_{\nu_3}^2 V_{\rm LM}^{3i\star} V_{\rm LM}^{3j}
\right)a_0 m_0
\log\frac{M_{\rm G}}{M_{\nu_R}},
\end{eqnarray}
with $i\ne j$. Here, we ignored the neutrino Yukawa coupling constants 
except for $f_{\nu_3}$ assuming they are small. 

In this model the LFV masses for the right-handed sleptons are negligibly 
small. This is an opposite case to the minimal SUSY SU(5) 
GUT  because the left-handed leptons have two types of the Yukawa 
interactions, $f_{l}$ and $f_{\nu}$, while the right-handed 
leptons have only $f_l$. In the minimal SUSY SU(5) GUT with the 
right-handed neutrinos \cite{HNY} or the SUSY SO(10) GUT \cite{BHS}, 
the LFV masses for 
both left-handed and right-handed sleptons may be induced radiatively.

\section{Branching ratio of $\mu\rightarrow \e \gamma$ in the MSSM} 

In the MSSM the event rate of $\mu\rightarrow \e \gamma$ is significantly 
enhanced proportional to square of $\tan\beta$ when it is much larger than 
one. This comes from a fact the MSSM has two Higgs doublet bosons. 
The process $\mu^+\rightarrow \e^+ \gamma$ 
is described by following effective electromagnetic-dipole type matrix element:
\begin{eqnarray}
  T=e \epsilon^{\alpha *}\bar{v}_\mu i \sigma_{\alpha \beta}
  q^\beta (A_L P_L + A_R P_R) v_{\e},
\label{Penguin}
\end{eqnarray}
where  $P_{R/L}=(1\pm\gamma_5)/2$, $q$ a photon momentum, and 
$\epsilon^{\alpha}$ is the photon polarization vector. From this equation, 
the event rate is given as
\begin{eqnarray}
\Gamma(\mu \rightarrow \e~\gamma)
= \frac{e^2}{16 \pi} m_{\mu}^3 (|A_L|^2+|A_R|^2).
\end{eqnarray}
The coefficients $A_R$ and $A_L$ have to be proportional to a Yukawa coupling
constant of lepton and one of the vacuum expectation value of the doublet
Higgs bosons, since the matrix element is violating both  
the SU(2)$_L\times$U(1)$_Y$ and the lepton chiral symmetries. Then, 
when $\tan\beta$ is large, the contribution to $A_L$ and $A_R$, proportional to 
$m_l\tan\beta (=f_l\langle h_f  \rangle)$, dominates over 
those proportional to 
$m_l(=f_l \langle \bar{h}_f \rangle)$.

As explained in the previous section, in the minimal SUSY SU(5) GUT, 
the right-handed sleptons have the LFV masses while the left-handed
sleptons do not. In this case, $A_R$ has a sizable contribution 
proportional to $m_\mu$, however $A_L$ not, and 
the diagrams contributing to $A_R$
interfere with each others destructively \cite{HMTY}. For simplicity, we explain it in a 
case with large $\tan\beta$.
In this case, two diagrams (Fig.~(1)) dominate over the other diagrams. 
The diagram (a) is proportional to $m_\mu\tan\beta$
through the left-right mixing mass term for slepton, while the diagram (b) 
is proportional to it through the Yukawa interaction of the Higgsino to 
slepton and the SU(2)$_L\times$U(1)$_Y$ breaking mixing mass between bino and   
the Higgsino. Assuming the minimal supergravity scenario, they are 
almost the same order of magnitude, and the relative phase between 
them is negative. Then, this distractive interfere reduces the event rate.
This destructive interference may also occur in small $\tan\beta$.

In the Fig.~(2) we show dependence of the branching ratio for $\mu 
\rightarrow \e \gamma$ on the right-handed selectron mass $m_{\tilde \e_R}$.
We choose the bino mass $M_1$ is 65 GeV, $a_0=0$, $\tan \beta=3$, 10, 30, and
the Higgsino mass $\mu>0$. Here, the top quark mass is 175GeV.  
Also, we impose the radiative
breaking condition of the SU(2)$_L\times$U(1)$_Y$ symmetry, and 
the experimental constraints including the anomalous magnetic dipole 
moment of muon. 
In this figure, there exists a region where the cancellation 
between the diagrams reduces the event rate of 
$\mu \rightarrow \e \gamma$ significantly. 

In the MSSM with the right-handed neutrinos, the left-handed sleptons may
have the LFV mass terms while the right-handed sleptons do not.
In this case, only $A_L$ gets the sizable contribution  proportional 
to $m_\mu$. Opposite to the minimal SUSY SU(5) GUT, a diagram 
dominates over the other diagrams \cite{HMTYY}. 
When $\tan\beta$ is large, Fig.~(3) is the dominant contribution.
Since we do not have enough information about the Yukawa coupling constants 
of neutrino to calculate the LFV event rate in detail,
we assume that the Yukawa coupling constant of tau neutrino is as large
as that of top quark, and the mixing matrix $V_{\rm LM}$ is given
by the CKM matrix. Also, we take $m_{\nu_\tau}=10$eV.
In the Fig.~(4) we show dependence of the branching 
ratio for $\mu \rightarrow \e \gamma$ on the left-handed selectron mass 
$m_{\tilde \e_L}$. We choose the wino mass $M_2$ is 130 GeV, $a_0=0$, $\tan \beta=3$, 
10, 30, and $\mu>0$, and the other gaugino masses are given by the 
GUT relation for simplicity. The branching ratio is enhanced by 
square of $\tan\beta$, and it can reach to one order of magnitude below
the experimental bound.

Finally, we discuss  $\mu \rightarrow \e \gamma$ in the SUSY SO(10) GUT.
Both the left-handed and right-handed sleptons may have the 
LFV mass terms in the SUSY SO(10) GUT. In this case, the dominant 
diagrams are proportional
to $m_\tau$, not $m_\mu$, and then, the branching ratio is enhanced by
about 100($\sim (m_\tau/m_\mu)^2$) \cite{BHS}. In large $\tan\beta$, 
Diagrams (a) and (b) in Fig.~(5) are dominant, and  contribute to 
$A_R$ and $A_L$, respectively.
Since the minimal SUSY SO(10) GUT can not induce the CKM matrix, 
we extend it to realize masses and mixing matrix of quarks and leptons.
In one of the models, the LFV mass terms of the left-handed and the 
right-handed sleptons can be given by the CKM matrix. We show the 
branching ratio of  $\mu \rightarrow \e \gamma$ in that case in Fig.~(6).
We choose the bino mass $M_1$ is 65 GeV, $a_0=0$, $\tan \beta=3$, 10, 30, and
$\mu>0$. The branching ratio can reach to the experimental upper bound.

Similar to the SUSY SO(10) GUT, if in the SUSY SU(5) GUT the extra 
interaction to $\Phi$ to realize masses and mixing of quarks and leptons
or the Yukawa coupling of the right-handed neutrinos is introduced,
it leads to sizable contribution to the LFV masses for the 
left-handed slepton, and the event rate of $\mu\rightarrow \e \gamma$ is also 
enhanced 
by $(m_\tau/m_\mu)^2$. For example, if we introduce the higher dimensional
interaction with the SU(5) breaking Higgs multiplet in order to explain
the mass ratio between down-type quarks and charged leptons, it may induce
the sizable LFV masses for the left-handed sleptons at large $\tan\beta$, 
and the 
event rate of  $\mu\rightarrow \e \gamma$ is significantly enhanced 
as Fig.~(6) \cite{HNOST}.

\section{Conclusion and Discussion}

We showed that, assuming the minimal supergravity scenario, we can 
probe the physics beyond the MSSM, the SUSY-GUT's, the seesaw mechanism 
with the right-handed neutrinos, and so on, through the LFV processes.
In the future experiments upper bounds for ${\rm Br}(\mu^+ \rightarrow \e^+ 
\gamma)$ 
is  expected to come down to $10^{-14}$ \cite{kuno,mega}.
While in the minimal SUSY SU(5) GUT the destructive interference
reduces the branching ratio significantly, the 
future experiments may be accessible to several models.

Finally, we discuss the other LFV processes. The experimental bound on 
the $\mu^--\e^-$ conversion rate in nuclei will be improved to the level of 
 $10^{-16}$ \cite{meco},
and it is also expected to give a severe constraint on the LFV physics.
In the MSSM, while penguin diagrams by photon and Z boson and box diagrams
contribute to it, the penguin diagrams induced by the photonic dipole term 
given 
by Eq.~(\ref{Penguin}) dominate over the other diagrams when $\tan\beta$
is large. In that case, the ratio of the conversion rate
and the branching ratio of $\mu\rightarrow \e\gamma$ is almost constant as
\begin{eqnarray}
{\rm R}(\mu^-\rightarrow \e^-;~^{48}_{22}{\rm Ti})&\simeq& 6 \times 10^{-3} 
{\rm Br}(\mu\rightarrow \e \gamma).
\end{eqnarray}
From this, it can understood that  the $\mu^--\e^-$ conversion in nuclei has 
sensitivity comparable to $\mu^+\rightarrow \e^+ \gamma$. Moreover, 
since the dependence of  $\mu^--\e^-$ conversion in nuclei on the SUSY 
parameters is different from that of $\mu\rightarrow \e\gamma$ when
 $\mu\rightarrow \e\gamma$ is suppressed by destructive interference,
searches for  $\mu^+\rightarrow \e^+\gamma$ and the $\mu^--\e^-$ conversion in 
nuclei are complementary.  

The dependence of the branching ratio of $\mu\rightarrow 3\e$ on the 
SUSY parameters tend to be similar to that of $\mu\rightarrow \e\gamma$
not only in large $\tan\beta$, but also small $\tan\beta$, since 
the phase space integral enhances the photonic penguin contribution.
The ratio between  the branching ratio $\mu\rightarrow 3\e$ 
and that of $\mu\rightarrow \e\gamma$ is given by
\begin{eqnarray}
{\rm Br}(\mu\rightarrow 3\e)&\simeq& 7 \times 10^{-3}  
                     {\rm Br}(\mu\rightarrow \e \gamma).
\end{eqnarray}
It is pointed out in Ref.~\cite{others} that if this process is discovered, 
we can study CP violation in the SUSY parameters.

The branching ratio of $\tau\rightarrow \mu\gamma$ will be improved
to the level of $10^{-(7-8)}$ in the B factories in KEK and  SLAC.  
The SUPERKAMIOKANDE experiment presents suggestive data that 
the atmospheric neutrino problem may come from the neutrino oscillation 
between $\nu_\tau$ and $\nu_\mu$ \cite{superkamiokande}. In that case,  if 
the tau neutrino Yukawa coupling constant is as large as that of top quark,
the branching ratio of $\tau\rightarrow \mu\gamma$ enter into a region
accessible to the experiments in the B factory \cite{HMTYY,HNY}. 

\newpage
%
%
\newcommand{\Journal}[4]{{\sl #1} {\bf #2} {(#3)} {#4}}
\newcommand{\APJ}{Ap. J.}
\newcommand{\CJP}{Can. J. Phys.}
\newcommand{\MPL}{Mod. Phys. Lett.}
\newcommand{\NC}{Nuovo Cimento}
\newcommand{\NP}{Nucl. Phys.}
\newcommand{\PL}{Phys. Lett.}
\newcommand{\PR}{Phys. Rev.}
\newcommand{\PRep}{Phys. Rep.}
\newcommand{\PRL}{Phys. Rev. Lett.}
\newcommand{\PTP}{Prog. Theor. Phys.}
\newcommand{\SJNP}{Sov. J. Nucl. Phys.}
\newcommand{\ZP}{Z. Phys.}

\newpage
%
%
%
%
\begin{figure}[p]
\begin{center} 
\begin{picture}(460,200)(0,0)
\ArrowLine(10,50)(45,50)    \Text(20,45)[t]{$\mu_L$}
\ArrowLine(220,50)(185,50)  \Text(200,45)[t]{${\rm e}_R$}
\DashArrowLine(45,50)(90,50){5} \Text(65,45)[t]{$\tilde{\mu}_L$}   
\DashArrowLine(140,50)(90,50){5} \Text(115,45)[t]{$\tilde{\mu}_R$}    
\DashArrowLine(185,50)(140,50){5} \Text(165,45)[t]{$\tilde{\rm e}_R$}    
\CArc(115,50)(70,0,180)\Text(115,140)[t]{$\tilde{B}^0$}
\Photon(180,105)(195,120){4}{4}  \Text(200,130)[l]{$\gamma$}
\Vertex(140,50){3}              
\Vertex(90,50){2} \Text(90,60)[b]{$\langle h_2 \rangle$}              
\Text(115,20)[b] {(a)}
\ArrowLine(275,50)(240,50)    \Text(250,45)[t]{$\mu_L$}
\ArrowLine(415,50)(450,50)  \Text(430,45)[t]{${\rm e}_R$}
\DashArrowLine(275,50)(345,50){5} \Text(310,45)[t]{$\tilde{\mu}_R$}   
\DashArrowLine(345,50)(415,50){5} \Text(385,45)[t]{$\tilde{\rm e}_R$}    
\CArc(345,50)(70,0,90)  \Text(345,140)[t]{$\langle h_2 \rangle$}
\CArc(345,50)(70,90,180)
\Text(305,130)[t]{$\tilde{h}$} \Text(385,130)[t]{$\tilde{B}^0$} 
\Photon(410,105)(425,120){4}{4}  \Text(430,130)[l]{$\gamma$}
\Vertex(345,50){3}              
\Vertex(345,120){2}              
\Text(345,20)[b] {(b)}
\end{picture}
\end{center}
\caption
{
Feynman diagrams giving dominant contribution to $\mu\rightarrow \e \gamma$
in large $\tan\beta$ when the right-handed sleptons have the LFV masses. The arrows represent
the chirality of lepton. 
}
\end{figure}
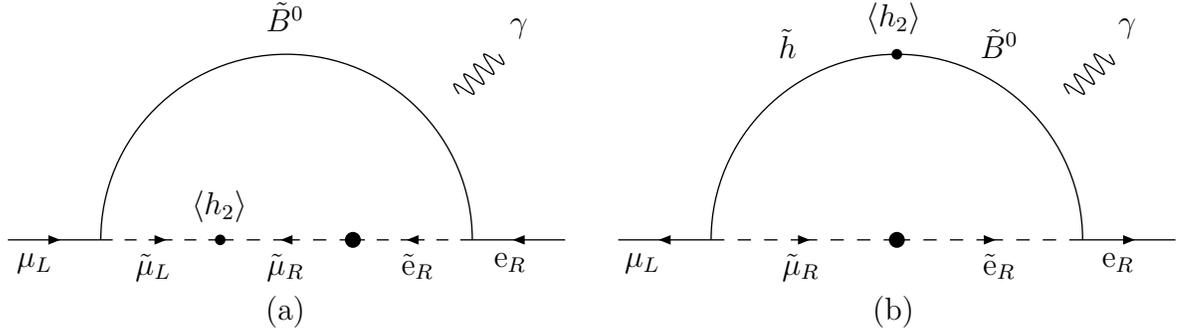
%
%
%
%
\begin{figure}[p]
\begin{center}
\leavevmode\psfig{figure=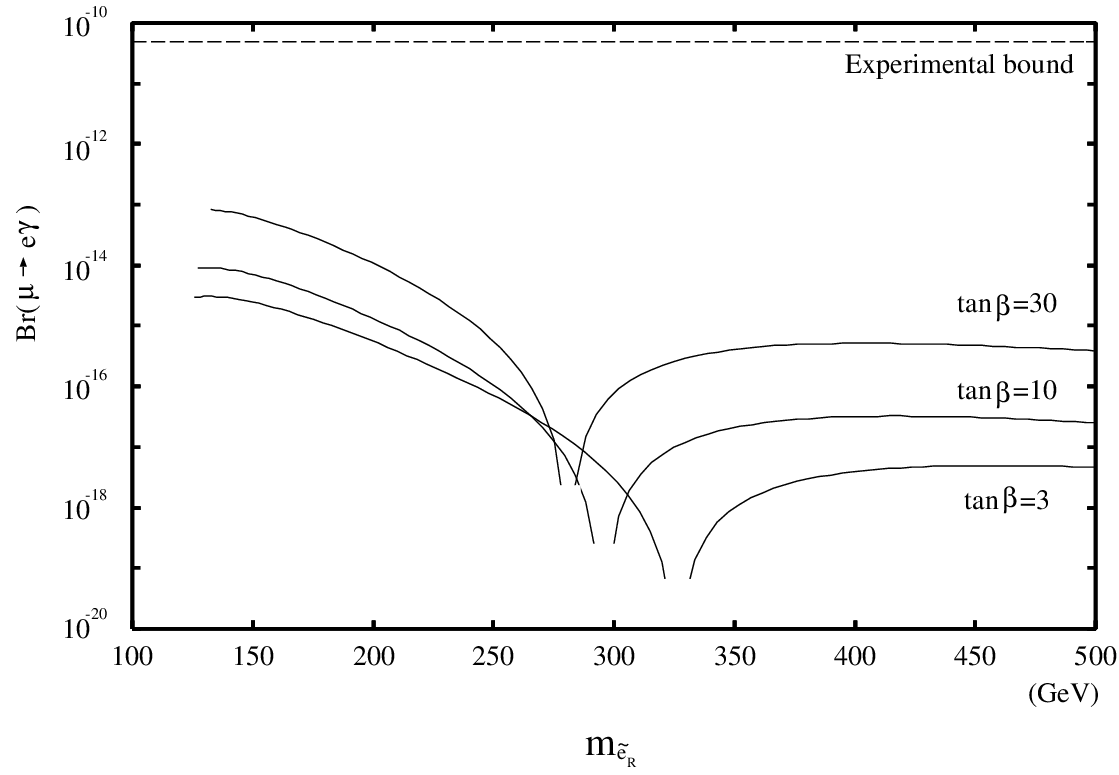,width=9cm}
\end{center}
\caption
{
The branching ratio of $\mu\rightarrow \e \gamma$ in the minimal SUSY SU(5) GUT
as a function of the physical right-handed selectron mass, $m_{{\tilde \e}_R}$.
Solid lines correspond to the cases for $\tan \beta=3,10,30$.
Dashed line represents the present experimental upper bound for this process. 
Here we take  the bino mass $M_1=65$ GeV, $a_0=0$, and the Higgsino mass $\mu > 0$.
}
\end{figure}
%
%
\begin{figure}
\begin{center} 
\begin{picture}(460,200)(0,0)
\ArrowLine(110,50)(145,50)    \Text(120,45)[t]{$\mu_R$}
\ArrowLine(320,50)(285,50)  \Text(300,45)[t]{${\rm e}_L$}
\DashArrowLine(215,50)(145,50){5} 
\Text(180,45)[t]{$\tilde{\mu}_L(\tilde{\nu}_\mu)$}   
\DashArrowLine(285,50)(215,50){5} 
\Text(255,45)[t]{$\tilde{\rm e}_L(\tilde{\nu}_{\rm e})$}    
\CArc(215,50)(70,0,90)  \Text(215,140)[t]{$\langle h_2 \rangle$}
\CArc(215,50)(70,90,180)
\Text(175,130)[t]{$\tilde{h}$} \Text(255,130)[t]{$\tilde{W}$} 
\Photon(280,105)(295,120){4}{4}  \Text(300,130)[l]{$\gamma$}
\Vertex(215,50){3}              
\Vertex(215,120){2}              
\end{picture}
\end{center}
\caption
{
Feynman diagrams giving dominant contribution to $\mu\rightarrow \e \gamma$
in large $\tan\beta$ when the left-handed sleptons have the LFV masses. The arrows represent
the chirality of lepton. 
}
\end{figure}
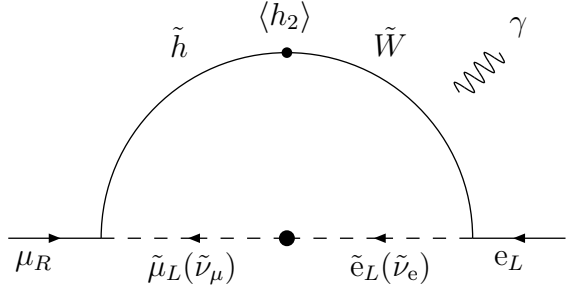
%
%
%
%
\begin{figure}[p]
\begin{center}
\leavevmode\psfig{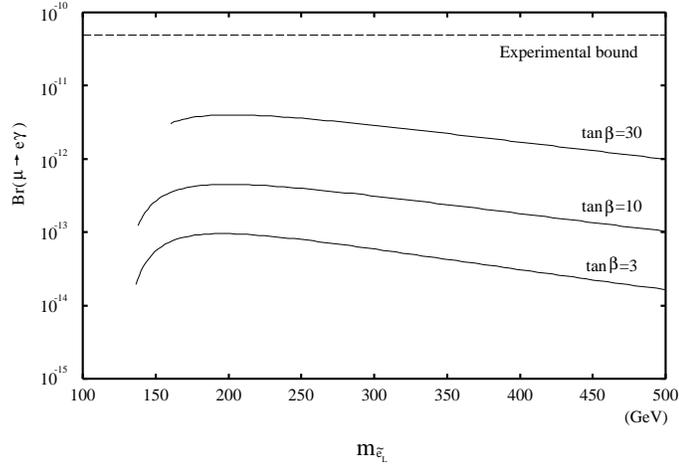}
\end{center}
\caption
{
The branching ratio of $\mu\rightarrow \e \gamma$ in the MSSM with the 
right-handed neutrinos as a function of the physical left-handed selectron 
mass, $m_{{\tilde \e}_L}$. We assume that the mixing matrix of lepton 
$V_{\rm LM}$ is given by the CKM matrix, and that the tau neutrino Yukawa 
coupling constant is as large as that of top quark.
We take $m_{\nu_\tau}=10$eV.
Solid lines correspond to the cases for $\tan \beta=3,10,30$.
Dashed line represents the present experimental upper bound for this process. 
Here, we take  the wino mass $M_2=130$ GeV, $a_0=0$, and the Higgsino mass $\mu > 0$.
}
\end{figure}
%
%
\begin{figure}
%
\begin{center} 
\begin{picture}(460,200)(0,0)
\ArrowLine(10,50)(45,50)    \Text(20,45)[t]{$\mu_L$}
\ArrowLine(220,50)(185,50)  \Text(200,45)[t]{${\rm e}_R$}
\DashArrowLine(45,50)(80,50){5} \Text(60,45)[t]{$\tilde{\mu}_L$}   
\DashArrowLine(80,50)(115,50){5} \Text(95,45)[t]{$\tilde{\tau}_L$}    
\DashArrowLine(150,50)(115,50){5} \Text(135,45)[t]{$\tilde{\tau}_R$}    
\DashArrowLine(185,50)(150,50){5} \Text(170,45)[t]{$\tilde{\rm e}_R$}    
\CArc(115,50)(70,0,180)\Text(115,140)[t]{$\tilde{B}^0$}
\Photon(180,105)(195,120){4}{4}  \Text(200,130)[l]{$\gamma$}
\Vertex(80,50){3}              
\Vertex(115,50){2} \Text(115,60)[b]{$\langle h_2 \rangle$}              
\Vertex(150,50){3}              
\Text(115,20)[b] {(a)}
\ArrowLine(275,50)(240,50)    \Text(250,45)[t]{$\mu_R$}
\ArrowLine(415,50)(450,50)  \Text(430,45)[t]{${\rm e}_L$}
\DashArrowLine(310,50)(275,50){5} \Text(290,45)[t]{$\tilde{\mu}_R$}   
\DashArrowLine(345,50)(310,50){5} \Text(325,45)[t]{$\tilde{\tau}_R$}    
\DashArrowLine(345,50)(380,50){5} \Text(365,45)[t]{$\tilde{\tau}_L$}    
\DashArrowLine(380,50)(415,50){5} \Text(400,45)[t]{$\tilde{\rm e}_L$}    
\CArc(345,50)(70,00,180)\Text(345,140)[t]{$\tilde{B}^0$}
\Photon(410,105)(425,120){4}{4}  \Text(430,130)[l]{$\gamma$}
\Vertex(310,50){3}              
\Vertex(345,50){2} \Text(345,60)[b]{$\langle h_2 \rangle$}              
\Vertex(380,50){3}              
\Text(345,20)[b] {(b)}
\end{picture}
\end{center}
\caption
{
Feynman diagrams giving dominant contribution to $\mu\rightarrow \e \gamma$
in large $\tan\beta$ when both the left-handed and right-handed sleptons 
have the LFV masses. The arrows represent the chirality of lepton. 
}
\end{figure}
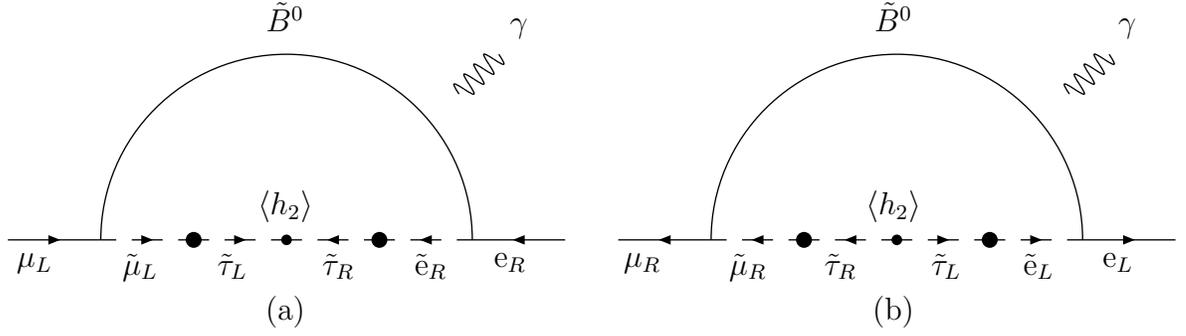

%
%
\begin{figure}[p]
\begin{center}
\leavevmode\psfig{figure=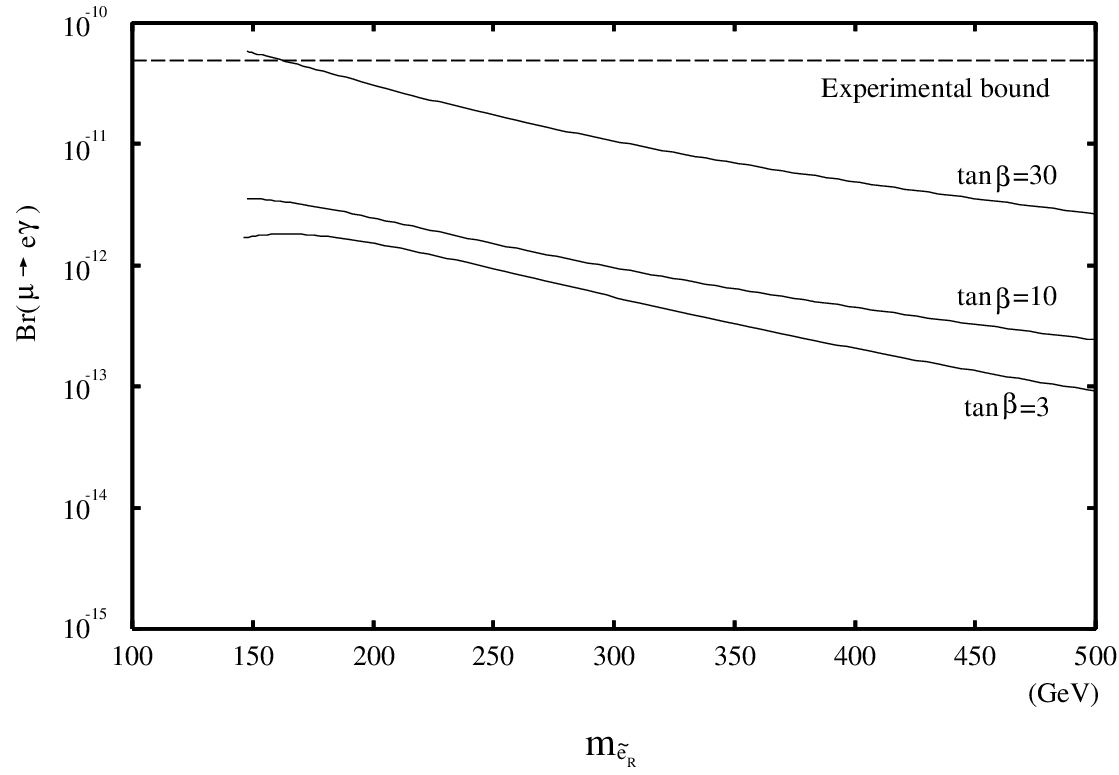,width=9cm}
\end{center}
\caption
{
The branching ratio of $\mu\rightarrow \e \gamma$ in the SUSY SU(10) GUT
as a function of the physical right-handed selectron mass, $m_{{\tilde \e}_R}$.
Solid lines correspond to the cases for $\tan \beta=3,10,30$.
Dashed line represents the present experimental upper bound for this process. 
Here we take  the bino mass $M_1=65$ GeV, $a_0=0$, and the Higgsino mass $\mu > 0$.
}
\end{figure}


\begin{thebibliography}{99}
%
\bibitem{EN}
        J. Ellis and D.V.~Nanopoulos, 
	            \Journal{\PL}{110B}{1982}{44};\\
        I-Hsiu~Lee, \Journal{\PL}{138B}{1984}{121};
                    \Journal{\NP}{B246}{1984}{120}.

\bibitem{nilles}
        For review, H.P.~Nilles,
        \Journal{\PRep}{110}{1984}{1}.
            
\bibitem{RG} 
	See review, G.F.~Giudice and  R.~Rattazzi,
		    CERN-TH-97-380 (hep-ph/9801271).
\bibitem{PL}
	For review, P.~Langacker, 
	            \Journal{\PRep}{72}{1981}{185}.
\bibitem{seesaw}
        T. Yanagida,
        in {\sl Proceedings of the Workshop on Unified Theory and
        Baryon Number of the Universe},
        eds. O. Sawada and A. Sugamoto (KEK, 1979) p.95; \\
        M. Gell-Mann, P. Ramond, and R. Slansky,
        in {\sl Supergravity},
        eds. P. van Nieuwenhuizen and D. Freedman
        (North Holland, Amsterdam, 1979).
	
\bibitem{HKR}
        L.~Hall, V.~Kostelecky, and S.~Raby, 
         \Journal{\NP}{B267}{1986}{415}.

\bibitem{DNNS}
        M.~Dine, A.E.~Nelson, Y.~Nir, and  Y.~Shirman,
	 \Journal{\PR}{D53}{1996}{2658}.

\bibitem{DS}
	S.~Dimopoulos and  D.~Sutter,
	\Journal{\NP}{B452}{1995}{496}.

\bibitem{BH}
        R.~Barbieri and L.~Hall,
        \Journal{\PL}{B338}{1994}{212}.

\bibitem{dimen5}
	N.~Sakai and T.~Yanagida, \Journal{\NP}{B197}{1982}{533};\\
	S.~Weinberg, \Journal{\PR}{D26}{1982}{287}.


\bibitem{proton}
	P.~Nath, A.~Chamseddine, and  R.~Arnowitt, 
		{\sl Phys. Rev.} {\bf D32} (1985) 2348;\\
	J.~Hisano, H.~Murayama, and T.~Yanagida, 
		\Journal{\PRL} {69}{1992}{1014};
		\Journal{\NP} {B402}{1993}{46};\\
	J.~Hisano, T.~Moroi, K.~Tobe, and T.~Yanagida,
		\Journal{\MPL}{A10}{1995}{2267}.
	
\bibitem{BM}
        F.~Borzumati and A.~Masiero,
        \Journal{\PRL}{57}{1986}{961}.

\bibitem{ACH}
        N.~Arkani-Hamed, H.~Cheng, and L.~Hall,
        \Journal{\PR}{D53}{1996}{413}.

\bibitem{HNY}
	J.~Hisano, D.~Nomura, and T.~Yanagida,
	KEK-TH-548 (hep-ph/9711348).

\bibitem{BHS}
        R.~Barbieri, L.~Hall, and A.~Strumia,
        \Journal{\NP}{B445}{1995}{219}.

\bibitem{HMTY}
        J.~Hisano, T.~Moroi, K.~Tobe, and M.~Yamaguchi,
        \Journal{\PL}{B391}{1997}{341};
        {\it Erratum-ibid}
        {\bf B397}{(1997)}{357}.

\bibitem{HMTYY}
	 J.~Hisano, T.~Moroi, K.~Tobe, M.~Yamaguchi, and T.~Yanagida,
	\Journal{\PL}{B357}{1995}{579};\\
       J.~Hisano, T.~Moroi, K.~Tobe, and M.~Yamaguchi,
        \Journal{\PR}{D53}{1996}{2442}.

\bibitem{HNOST}
	J.~Hisano, D.~Nomura, Y.~Okada, Y.~Shimizu, and M.~Tanaka,
	KEK-TH-575 (hep-ph/9805367).

\bibitem{kuno}
	Y.~Kuno and Y.~Okada,
        	\Journal{\PRL}{77}{1996}{434};\\
	Y.~Kuno, A.~Maki, and Y.~Okada,
        	\Journal{\PR}{D55}{1997}{2517};\\
	Y.~Kuno, KEK-Preprint-97-59.
%
\bibitem{mega}
	M.D.~Cooper, Talk on the fourth KEK typical conference 
	(Tsukuba, Oct, 1996).
%
\bibitem{meco}
	MECO collaboration, Proposal to Brookhaven National Laboratory AGS 
	(Sep, 1997).
%
\bibitem{others}
	Y.~Okada, K.~Okumura, and Y.~Shimizu,
	 KEK-TH-535 (hep-ph/9708446).

\bibitem{superkamiokande}
Super-Kamiokande Collaboration,
ICRR-REPORT-418-98 (hep-ex/9805006).

\end{thebibliography}
\end{document}